\documentclass[11pt,a4paper]{article}
\usepackage{amsmath,amssymb,amsfonts,slashed}
\usepackage[text={176mm,240mm},centering]{geometry}
\usepackage{float}

\usepackage{longtable}
\usepackage[titletoc]{appendix}
\usepackage[figuresright]{rotating}
\usepackage{multirow}
\usepackage{array,multirow,makecell}
\usepackage{graphicx}
\usepackage{color}
\usepackage{gensymb}
\usepackage[mathscr]{euscript}
\usepackage{array,multirow,makecell}
\usepackage{booktabs}
\usepackage[
colorlinks=true,
linkcolor=red,
breaklinks=true,
urlcolor=blue,
citecolor=blue]{hyperref}

\newcommand \bra [1] {\langle {#1}\vert}
\newcommand \ket [1] {\vert {#1}\rangle}

\def\be{\begin{equation}}
\def\ee{\end{equation}}
\def\ba{\begin{eqnarray}}
\def\ea{\end{eqnarray}}
\def\3P0{{}^3P_0}
\newcommand \helicityAmp {\mathcal{M}^{m_{J_A}m_{J_B}m_{J_C}}}

\begin{document}
\title{Modeling Charmonium-$\eta$ Decays of $J^{PC}=1^{--}$ Higher Charmonia}

\date{\today}

\author{
     Muhammad Naeem Anwar$^{1,2}$
                          \footnote{Email address:\texttt{naeem@itp.ac.cn}}~,
     Yu Lu$^{2,3}$
                        \footnote{Email address:\texttt{luyu211@ihep.ac.cn}}~,
     Bing-Song Zou$^{1,2}$
                        \footnote{Email address:\texttt{zoubs@itp.ac.cn}}
       \\[2mm]
      {\it\small$^1$CAS Key Laboratory of Theoretical Physics, Institute of Theoretical Physics,}\\
      {\it\small Chinese Academy of Sciences, Beijing 100190, China}\\
      {\it\small$^2$University of Chinese Academy of Sciences, Beijing 100049, China}\\
      {\it\small$^3$Institute of High Energy Physics, Chinese Academy of Sciences, Beijing 100049, China}\\
}

\medskip
\maketitle

\begin{abstract}
We propose a new model to create a light meson in the heavy quarkonium transition, which is inspired by the Nambu$-$Jona-Lasinio (NJL) model. Hadronic transitions of $J^{PC}=1^{--}$ higher charmonia with the emission of an $\eta$ meson are studied in the framework of the proposed model. The model shows its potential to reproduce the observed decay widths and make predictions for the unobserved channels. We present our predictions for the decay width of $\Psi \to J/\psi \eta$ and $\Psi \to h_{c}(1P)\eta$, where $\Psi$ are higher $S$ and $D$ wave vector charmonia, which provide useful references to search for higher charmonia and determine their properties in forthcoming experiments. The predicted branching fraction $\mathcal B(\psi(4415)\to h_{c}(1P)\eta)=4.62\times 10^{-4}$ is one order of magnitude smaller than the $J/\psi \eta$ channel. Estimates of partial decay width $\Gamma(Y \to J/\psi \eta)$ are given for $Y(4360)$, $Y(4390)$ and $Y(4660)$ by assuming them as $c\bar{c}$ bound states with quantum numbers $3 ^3D_1$, $3 ^3D_1$ and $5 ^3S_1$, respectively. Our results are in favor of these assignments for $Y(4360)$ and $Y(4660)$. The corresponding experimental data for these $Y$ states has large statistical errors which do not provide any constraint on the mixing angle if we introduce $S-D$ mixing. To identify $Y(4390)$, precise measurements on its hadronic branching fraction are required which are eagerly awaited from BESIII.
\end{abstract}
\thispagestyle{empty}

\section{Introduction}

\begin{figure}[H]
  \centering
  \includegraphics[width=0.6\textwidth]{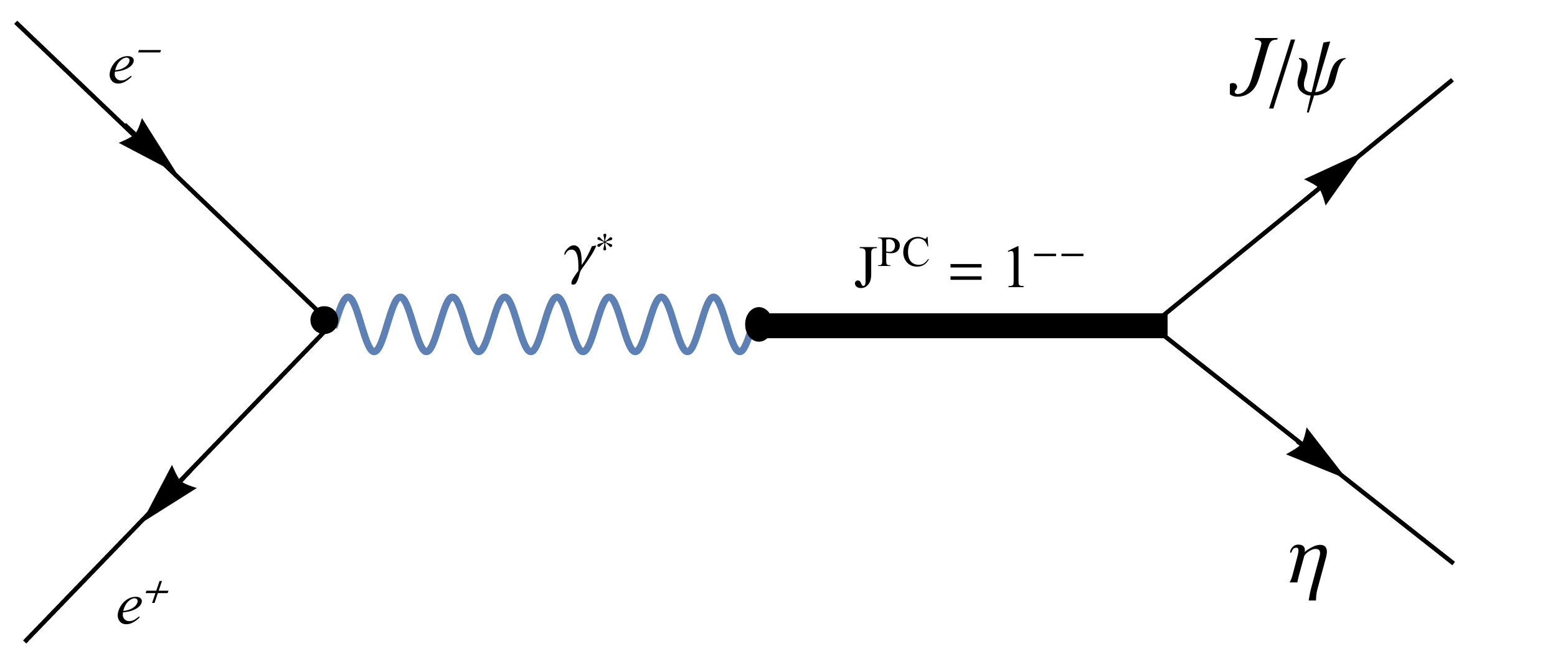}\\
  \caption{Feynman diagram associated with the intermediate production of higher vector charmonia via $e^{+}e^{-}$ annihilation, which further decay into $J/\psi \eta$.}
 \label{fyn1}
\end{figure}

Quantum chromodynamics (QCD), the gauge theory of strong interaction, received huge devolvement during the last few decades. However, it is still a subject of intensive research of various theoretical constructs (for instance, see recent review \cite{Brambilla:2014jmp}). Study of heavy quarkonium decays is a good probe to understand the nonperturbative nature of QCD at different energy regimes. Thanks to the wealth of experimental facilities like CLEO, Belle, \textit{BABAR}, CDF, D$0$, and BESIII, now we have intensive experimental data in the charmonium ($c\bar{c}$) energy regime. This provides us with great opportunities to test and explore the nature of strong interactions in the heavy quark sector. Currently BESIII is taking data in the $c\bar{c}$ energy regime and it is easy to produce $J^{PC}=1^{--}$ higher charmonia through $e^{+}e^{-}$ annihilations. Figure~\ref{fyn1} is the sketch of the intermediate production of vector charmonia, which further decay into $J/\psi \eta$. Since the center-of-mass energy of BESIII can go up to $4.6$ GeV~\cite{Asner:2008nq}, which is around the mass region of $\psi(5S)$ and $\psi(4D)$, it is a good opportunity to study the production and decay mechanism of higher vector charmonia. In the future $\bar{\textrm{P}}$ANDA also plans to collect data~\cite{Lutz:2009ff} in the $c\bar{c}$ energy regime which $e^{+}e^{-}$ colliders are not capable of producing directly. These experimental facilities will surely help us to deepen our understanding of heavy quarkonium physics and hence the nonperturbative aspects of strong interaction.

Heavy hadron spectroscopy has celebrated almost four decades since the discovery of $J/\psi$ in 1974. During this era, many theoretical studies have been carried out in the quark model framework to produce the spectrum of heavy quarkonium systems~\cite{Eichten:2007qx,Brambilla:2010cs}. An important manifestation of studying heavy $Q\bar{Q}$ is that the spectrum of these states can be explained by using nonrelativistic formalism. For instance, the Cornell potential model \cite{Eichten:1978tg,Eichten:1979ms}, which incorporates a spin-independent color Coulomb plus scalar linear confined potential, was hugely successful at describing the spectrum of the charmonium systems. The Cornell spin-independent potential is an approximate heavy quark spin symmetry (HQSS) within $c\bar{c}$ systems. Deviations with experiments can be observed in such potential models due to HQSS breaking effects~\cite{Brambilla:2004wf,Cao:2016xqo}. One possible source of breaking HQSS is the spin-dependent potential which introduces relativistic corrections to the Cornell potential model. The widely used relativized quark potential model, sometimes also referred as the Godfrey-Isgur (GI) model~\cite{Godfrey:1985xj}, is so far considered the best available option to reproduce the spectra of heavy quarkonium systems.

In heavy $Q\bar{Q}$ systems, hadronic transitions serve as a crucial probe of their internal structures and help to establish the understanding of light quark coupling with a heavy degree of freedom. In QCD, the well-established formalism for hadronic transitions is multipole expansion (ME) \cite{Gottfried:1977gp,Voloshin:1978hc,Yan:1980uh,Kuang:2006me}, which assumes that these transitions take place due to the intermediate process of gluon emission. These gluons are supposed to be soft, having wavelengths much larger than the size of a heavy quarkonium system. These soft gluons further couple to $\pi$(s) and $\eta$ to complete such kinds of hadronic transitions.

Development of heavy meson chiral Lagrangians (HMCL)~\cite{Casalbuoni:1996pg} is the foremost simplification to QCDME. HMCL serve as an effective field theory (EFT) to QCDME in a soft exchange where the gluonic exchanges are predominantly of limited momenta. With the assumptions that (i) the heavy $Q\bar{Q}$ involved in the process is well separated to consider it in a stringlike picture and (ii) the momentum of the emitted light meson is not too large, the HMCL are successful at reproducing the hadronic transitions among lower charmonia~\cite{Casalbuoni:1992yd,Casalbuoni:1992fd}. The experimental status of the spectrum of higher vector charmonium(like) states is very rich now and several precise measurements have been recorded for their hadronic transitions~\cite{Olive:2016xmw}. To describe the observed transitions of higher $c\bar{c}$ systems there is a potential need for a theoretical model which can predict the transitions in the high momentum regime and help to identify the missing higher states through their hidden-flavor decays. We try to fulfill this need by modeling hadronic transitions of higher vector charmonia. Our proposed model is away from all the assumptions [(i) and (ii)] of HMCL and useful to predict the transitions involving much large momenta.

Another possibility is that the transition between two $S$ waves, $S$ to $P$ or $D$ to $S$ wave charmonia with the emission of $\eta$ ($\pi$) might occur through intermediate open-charm contributions. Heavy quarkonium states can couple to intermediate heavy mesons through the creation of a light quark-antiquark pair. The formalism which incorporate intermediate heavy mesons within hadrons is sometimes referred to as coupled-channel effects. For instance, using the $^3P_0$ quark pair creation mechanism~\cite{Micu:1968mk}, intermediate meson loop contributions are found to be essential to explain the suppression of dielectric decay widths of higher bottomonium~\cite{Lu:2016mbb}. Coupled-channel effects have also been taken into account in the QCDME framework to study the hadronic transitions with two-pion emission for the charmonium system and found a good agreement with experimental measurements \cite{Zhou:1990ik}. In this paper, we neglect the coupled-channel effects for simplicity, which can be included in the future in the unquenched quark model.

To investigate the intermediate charmed meson loop effects on $\psi(3686) \to J/\psi \eta$ decay, nonrelativistic effective field theory (NREFT) formalism was constructed~\cite{Cao:2016xqo,Guo:2009wr,Guo:2010ak}. It is noted that if we go to much higher waves e.g., $\psi(nS)$ or $\psi((n-1)D)$ with $n=4,5,6...$, the decay momentum is not so small, as it lies in the relativistic regime; hence, the NREFT formalism is not very suitable for studying hadronic transitions of higher charmonia.

These indications bring out the fact that there is a need for a comprehensive model which is capable of producing hadronic transitions with the emission of light meson(s) for higher mass charmonium systems. We attempt to fill this gap by modeling the transitions $\Psi\to J/\psi \eta$ and $\Psi\to h_{c}(1P)\eta$, where $\Psi$ refers to $n ^3S_1$ and $(n-1) ^3D_1$ vector charmonia with ($n=2,3,4,...$). We present our predictions for hadronic transitions of higher vector charmonia into $J/\psi \eta$ and $h_{c}(1P)\eta$, which provide useful references to determine their properties in ongoing and forthcoming experiments.

In hadron physics, the most widely used model to study open-flavor strong decays is $^3P_0$ or the quark pair creation (QPC) model. Within the framework of the $^3P_0$ model, quark pair creation is induced from QCD vacuum. Hence, the generated quark pair shares the quantum numbers of vacuum ($J^{PC}=0^{++}$); therefore, it is referred to as the $^{3}P_{0}$ pair creation mechanism. The traditional $^3P_0$ model has been widely used in hadron spectroscopy and decays \cite{Lu:2016mbb,Ackleh:1996yt,Barnes:2005pb,Ferretti:2014xqa}. In the $^3P_0$ model, the probability to generate $q\bar{q}$ pairs is independent of the distance of the generation point from the valence quarks. In this work, we introduce the pair creation triggered from the Nambu$-$Jona-Lasinio (NJL) four-point-like effective interaction ($\mathcal{L}_{\textrm{NJL}}$). Since $\mathcal{L}_{\textrm{NJL}}$ is a mixture of scalar and pseudoscalar interactions, it raises the quantum numbers of the created $q\bar{q}$ pair as a mixture of $^{3}P_{0}$ and $^{1}S_{0}$. It should be noted that the dynamics of the creation of a $^{3}P_{0}$ vertex in the NJL framework is totally different from the conventional $^{3}P_{0}$ mechanism.

Recently, the interaction of $\mathcal{L}_{\textrm{NJL}}$ has been used to study the mixing of $\Omega (sss)$ baryons with its pentaquark partner states and it is found that the $\mathcal{L}_{\textrm{NJL}}$ leads to strong mixing between three-quark and five-quark $sss\leftrightarrow sssq\bar{q}$ (where $q=u,d,s$), with $J^P=\frac{3}{2} ^-$. It was reported that this expected mixing results in the decrease of the energy of the lowest $\Omega$ state \cite{An:2014lga}. There is no hint of such kind of mixing within the conventional $^{3}P_{0}$ model (which only involves the scalar interaction). Hence these are charming motivations to consider this interaction to study the hadronic decays of higher quarkonia.

The paper is organized as follows. In Sec.~\ref{model}, we review the development of the NJL model and its application in hadron spectroscopy and decays, where we deduce the effective Lagrangian for hadronic transitions with the emission of light meson(s). We conclude this section with an overview of well-established $S-D$ mixing formalism. Section~\ref{results} is devoted to discussing the results for $\Gamma(\Psi \to J/\psi \eta)$, $\Gamma(\Psi \to h_{c}(1P) \eta)$, and estimates of the branching fraction $\mathcal B(Y\to J\psi \eta)$ for $Y(4360)$, $Y(4390)$, and $Y(4660)$. Finally, we give a short summary in Sec.~\ref{summary}.

\section{Theoretical Framework}
\label{model}
\subsection{NJL Motivated Effective Lagrangian}
\label{NJL}

Effective field theories are very useful when the dynamics of the system involves only a few relevant degrees of freedom instead of all. The NJL model is one of the best examples of such kinds of effective theories which have the capability to recover almost all of the features of the exact leading theory. Historically, Nambu and Jona-Lasinio modeled a scheme to explain the pions as nucleon-antinucleon bound states \cite{Nambu:1961tp,Nambu:1961fr}; afterwards, the scheme gained more appreciation for being used at a more microscopic level by changing the nucleon field into a quark field $\psi$. NJL model only involves the quark degree of freedom, while the gluon degree of freedom is frozen in its point-like interaction vertex. NJL four-point-like interaction between quarks can be described by
\begin{equation}
\mathcal{L}_{\textrm{NJL}}=\frac{1}{2}g_{s}\sum_{a=0}^{N_{c}}\big[(\bar{\psi}\lambda^{a}\psi)^{2}+(\bar{\psi}\lambda^{a}i\gamma^{5}\psi)^{2}\big],
\end{equation}
where $N_{c}=8$ indicates the color degree of freedom; $\lambda^{a} (a=1,\cdot\cdot\cdot,8)$ are Gell-Mann matrices in $SU(3)$ flavor space with flavor singlet $\lambda^{0}=\sqrt{\frac{2}{3}}\mathcal I$, where $\mathcal I$ is the unit matrix in the three-dimensional flavor space, $\psi$ represents the quark field and $g_{s}$ is the coupling strength. This four-point-like color interaction with only one free parameter ($g_{s}$) has the capability to produce quite good results for the spectrum of low-lying and excited light mesons \cite{Klimt:1989pm,Vogl:1989ea}. Using the NJL motivated SU(2)$\otimes$SU(2) chiral Lagrangian for the excited pions, $\rho$ and $\omega$ mesons, the strong decay widths for the $V'\rightarrow PP$, $V'\rightarrow VP$, $P'\rightarrow VP$ transitions ($V'$ and $P'$ are the excited vector and pseudoscalar meson decaying into the vector $V$ and pseudoscalar $P$ meson, respectively) have been computed and found to be in good agreement with the experimental data \cite{Ebert:1994mf,Volkov:1998fa}. During the $1990$s, attempts were made to extend this approach to study the radiative transitions and strong decays of the charmed mesons. It is noted that the qualitative estimates of the strong decay widths using this approach agree well with the experimental data \cite{Khanna:1993xx}.

\begin{figure}[H]
  \centering
  \includegraphics[width=0.6\textwidth]{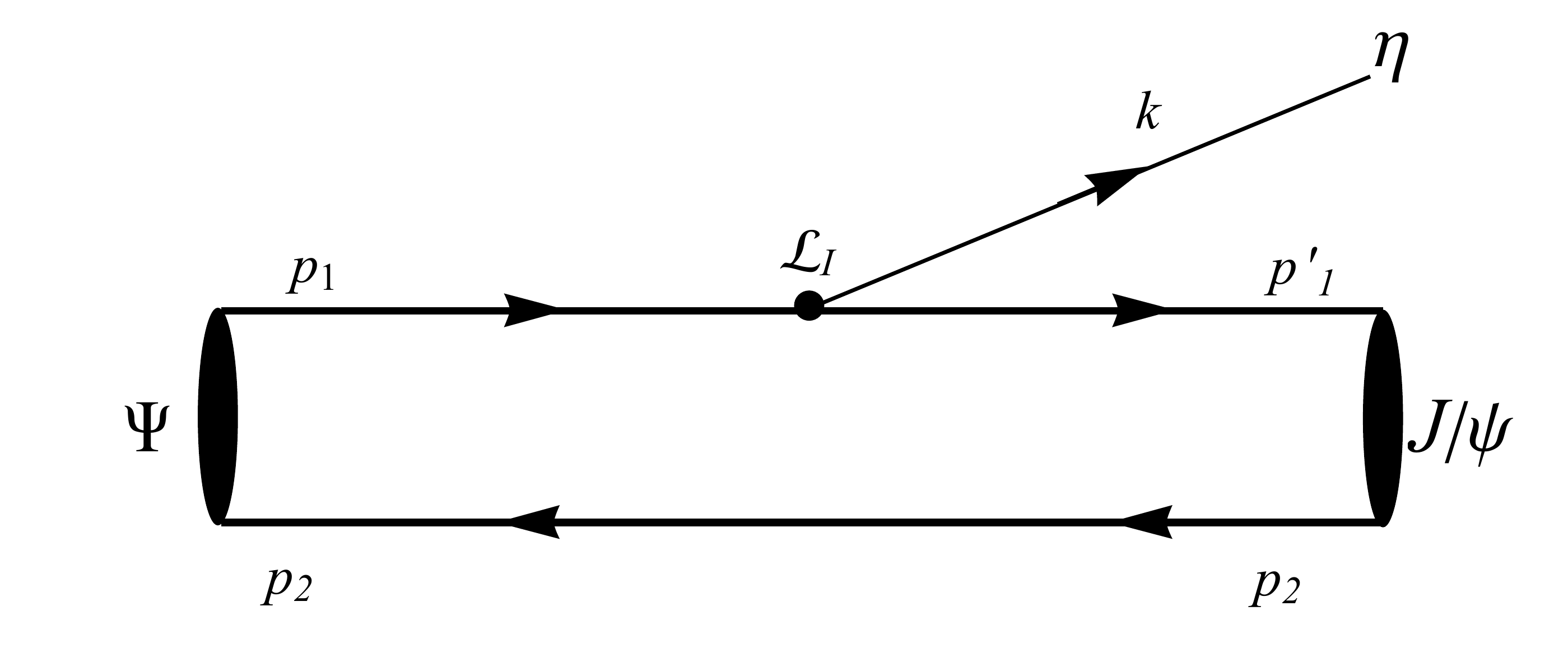}\\
  \caption{Quark level diagram of higher vector charmonia decaying into $J/\psi \eta$.}
 \label{fyn2}
\end{figure}

In light of these phenomenological studies we propose another quark pair creation mechanism which is inspired by the NJL model. We model the coupling of the light scalar and pseudoscalar meson with the charm quark. The effective Lagrangian of our model contains both the scalar and pseudoscalar interactions as present in the NJL model. In the case when we link the light $q\bar{q}$ production with an (anti)quark line, the effective Lagrangian of our proposed model can be written as
\be
\mathcal{L}_I=g(\bar{\psi} \psi <\sigma>+\bar{\psi} i\gamma^5 \psi <\eta>),
\ee
where $g$ is the overall coupling strength, $\psi$ is the heavy quark field, and $<\sigma>$ and $<\eta>$ are $SU(3)$ singlet scalar and pseudoscalar meson, respectively. Since $\bar{\psi}\psi$ is the $SU(3)$ singlet, the light sector should also be in a singlet. That is why the above Lagrangian does not have $SU(3)$ flavor matrices as present in $\mathcal{L}_{\textrm{NJL}}$. The color index can also be suppressed. The above Lagrangian $\mathcal{L}_I$ allows the coupling of the (anti)quark line only to a scalar or isospin singlet pseudoscalar. The possible Feynman diagram for the process $\Psi \to J/\psi \eta$ is shown in Fig.~\ref{fyn2}.

In principle, $\eta$ can also couple to the antiquark line, so one needs to calculate two diagrams. But going through the details, it becomes clear that both diagrams are equally contributing.

Since the experimental data are available for $\Psi \to J/\psi \eta$, it provides us with an opportunity to test this model. No experimental data are available for the emission of a light scalar meson, so we mainly focus on $\eta$ emission transitions to check the validity of the model. It is quite possible that our proposed model can be extended to produce di-pion transitions. It would be through the intermediate production of a $\sigma$ meson, which further decays into $\pi^+ \pi^-$. But it will involve final state interactions (FSI) between $\pi^+$ and $\pi^-$ as investigated in Ref.~\cite{Blundell:1996as}. This would be interesting but leads to intensive work which we will consider in our future studies.

To calculate the hadronic matrix elements, we prefer the mock hadron prescription~\cite{Hayne:1981zy} to express the initial and final meson wave functions. The mock hadron is defined as a collection of free quarks with the wave function of the bound quarks in a physical hadron normalized to the physical mass. The advantage of using this prescription is that it lets us calculate hadronic amplitudes as integrals over free-quark amplitudes. The initial mock meson wave function can be expressed as
\be
\vert A\rangle= \sqrt{2 E_A} \sum_{LS} \langle Lm, S S_z\vert J_A m_{JA}\rangle \int d^3 p_1 \phi_A(p_1) \chi_A^{12} \vert q_1(p_1),\bar{q}_2(-p_1)\rangle,
\ee
where $E_A$ is the total energy of the meson; $\vert Lm\rangle$, $\vert S S_z\rangle$, and $\vert J_A m_{JA}\rangle$ are the orbital angular momentum between the quark and antiquark, the total spin of the quark-antiquark pair, and the total angular momentum of the meson, respectively; $\langle Lm, S S_z\vert J_A m_{JA}\rangle$ is the Clebsch-Gordan coefficient; and $\phi_A$ and $\chi_A^{12}$ are the spatial and spin wave functions of the initial meson $\vert A\rangle$, respectively. The relative momentum between the quark and antiquark, $p_{1}$, is integrated over all values. We adopt the relativistic normalization
\be
\langle A(p')\vert A(p) \rangle=2 E_A \delta^3(\vec{p}-\vec{p}\,').
\ee
For two-body $A \to BC$ decay, we define the transition amplitude as
\be
\bra{BC} \mathcal H_I \ket{A} = 2\pi \sqrt{8E_A E_B E_C}\delta^4(p_i-p_f) \helicityAmp.
\ee
Considering the standard relativistic phase space, we define the decay width in the center-of-mass (CM) frame as
\be
\Gamma_{A\to BC}=2 \pi k \frac{E_B E_C}{m_A} \sum_{m_{J_B},m_{J_C}} \int d\Omega_B\vert \helicityAmp \vert^2,
\label{width}
\ee
for any fixed $m_{J_A}$. Since the decay width is independent of the polarization of the initial state, we set $m_{J_A}=J_A$ in the following calculations. Here $k$ expresses the momentum of the outgoing mesons $B$ or $C$, which is given as
\be
k=\frac{\sqrt{[m_{A}^2-(m_B-m_C)^2][m_{A}^2-(m_B+m_C)^2]}}{2m_A},
\label{mom}
\ee
with $E_{B}=\sqrt{m_B^2+k^2}$ and $E_{C}=\sqrt{m_C^2+k^2}$. The overlap of the wave functions of the initial meson $\ket A$ and the final mesons $\ket {B,C}$ can be expressed as
\be
\helicityAmp =g \int d^3 p_1
\phi_A(\vec{p}_1) \phi_B^*(\vec{p}_1-x_B \vec{P}_B) \mathcal{M}_0,
\label{amp}
\ee
where $x_B=m_{Q}/(m_Q+m_{\bar{Q}})=1/2$, with $m_Q=m_{\bar{Q}}=m_c$, with $m_c$ the charm quark mass. $\mathcal{M}_0$ is the free-quark amplitude which is discussed in detail in the Appendix~\ref{app:amp}. To find the overlap of the wave functions, we use simple harmonic oscillator (SHO) wave functions, which can be written in momentum space as
\be
\psi_{n_{r}lm}(\vec{p})=R_{n_{r}l}(p) \mathcal Y^{m}_{l}(p,\theta,\varphi),
\ee
where $n_{r}$, $l$, and $m$ represent the radial, orbital, and magnetic quantum numbers, respectively. $\mathcal Y^{m}_{l}(p,\theta,\varphi)=p^l Y^{m}_{l}(\theta,\varphi)$ is the solid harmonic defined as a function of spherical harmonic. The radial wave function $R_{n_{r}l}(p)$ is given as
\be
R_{n_{r}l}(p)= \sqrt{\frac{2n_{r}!}{\Gamma(n_{r}+l+\frac{3}{2})}}  \beta^{-\big(l +\frac{3}{2}\big)} e^{-p^{2}/2\beta^2} L_{n_{r}}^{l +\frac{1}{2}}(p^{2}/\beta^2),
\ee
where $\beta$ is an oscillatory parameter and $L_{n}^{l +\frac{1}{2}}(p^{2}/\beta^2)$ is the associated Laguerre polynomial. Indeed, SHO wave functions serve as a coarse approximation to the true wave functions. However, qualitatively, SHO wave functions are similar to the realistic wave functions and useful for producing analytical results.

\subsection{$S-D$ Mixing}
\label{mixing}

It is predicted that the $J^{PC} = 1^{--}$ charmonia near or above the open-charm threshold are an admixture of $S$ and $D$ waves~\cite{Ding:1991vu,Rosner:2001nm,Badalian:2008dv}. An $S$ wave dominant state has a $D$ wave component in its wave function and vice versa. A well-established formalism of this $S-D$ mixing is based on reproducing the dielectric decay widths to deduce the mixing angle. If we neglect the open-charm contributions due to coupling to corresponding decay channels, under the assumption that the $n ^3S_1$ state only mixes with $(n-1) ^3D_1$, $S$ and $D$ wave dominant states can be expressed as
\be
\psi_{\textrm{phys}}=\cos \theta \vert n ^3S_1 \rangle +\sin \theta \vert (n-1) ^3D_1 \rangle
\ee
\be
\psi'_{\textrm{phys}}=-\sin \theta \vert n ^3S_1 \rangle +\cos \theta \vert (n-1) ^3D_1 \rangle
\ee
Here $\psi_{\textrm{phys}}$ and $\psi'_{\textrm{phys}}$ represent the $S$ wave and $D$ wave dominant state, respectively. The relative sign between $\vert n ^3S_1 \rangle$ and $\vert (n-1) ^3D_1 \rangle$ is just a matter of convention. One may follow the other convention as in Ref. \cite{Rosner:2001nm}, but the effect of relative sign can be compensated by swapping $\theta\to -\theta$. A rough estimate of the $S-D$ mixing angle can be made by computing the ratio of the dielectric decay widths~\cite{Ding:1991vu,Rosner:2001nm}. This has been done in the quark model framework by computing the wave functions using a potential model and then tuning the mixing angle to reproduce the dielectric decay widths. Considering $\psi(3770)$ as the $1D$ dominant state with small $2S$ component, there exist two sets of possible ranges for the values of the mixing angle: $\theta\approx-10\degree \sim-13\degree$ and $\theta\approx+26\degree\sim+30\degree$ \cite{Ding:1991vu,Rosner:2001nm,Kuang:1989ub}.

There also exist a couple of quark-model-based phenomenological studies in favor of large $S-D$ mixing~\cite{Badalian:2008dv,Liu:2004un}. A large mixing angle such as $\theta = 34\degree$ is used~\cite{Badalian:2008dv} to produce almost the same dielectric decay widths of $\psi(4040)$ and $\psi(4160)$, which is consistent with experimental measurements~\cite{Olive:2016xmw}. The experimental fact that the $n ^3S_1$ dominant states have relatively small dielectric decay widths while $n ^3D_1$ dominant states have rather large widths can also be well described by considering the large $S-D$ mixing as the underlying mechanism~\cite{Badalian:2008dv}.

\section{Results and Discussions}
\label{results}
\subsection{$\Gamma(\Psi\to J/\psi \eta)$}
\label{widths}

To compute the decay widths for the process $\Psi\to J/\psi \eta$, it would be better to analyze the dependence of the wave function on the oscillatory parameter $\beta$ in the case of SHO wave functions. For light $q\bar{q}$ systems, the best value of $\beta$ from spectroscopy and decays is $0.379$\footnote{$\beta$ is in units of GeV; however, for the sake of simplicity, only its numerical values are written.} \cite{Ackleh:1996yt,Barnes:1996ff}. But for heavy $Q\bar{Q}$ states it is bit larger and also has a range $\beta=0.4\sim0.6$ in the literature \cite{Barnes:2005pb,Kalashnikova:2005ui,Liu:2011yp,LeYaouanc:1977fsz,LeYaouanc:1977gm}. The parameter $\beta$ relates to the size of quark-antiquark bound state. Since the size of heavy $Q\bar{Q}$ is smaller than the light $q\bar{q}$ system, $\beta_{q\bar{q}}\sim \Lambda_{\textrm{QCD}}$ for $q=u,d,s$ and $\beta_{Q\bar{Q}}\sim m_{Q}v$ for $Q=c,b$, where $m_Q$ is the mass and $v$ is the velocity of the heavy quark. To choose the same $\beta$ for initial and final mesons is quite reasonable~\cite{Blundell:1996as}, although there exist some predictions that each meson has its own effective $\beta$~\cite{Kokoski:1985is}. Quark model studies show that the effective $\beta$ for higher $c\bar{c}$ multiplets is smaller than the corresponding lower ones. It is due to the fact that the excited states have large spatial extensions. For instance, for $1P$ multiplets, $\beta=0.514$ and for $2P$ multiplets, it is $0.435$ \cite{Kalashnikova:2005ui}. This is an indication that for higher charmonium states, the favorable value of $\beta$ should be around $0.4$. The value $\beta =0.44$ has also been used to incorporate the spin counting predictions of open-flavor strong decays of higher $S$ wave $c\bar{c}$ states under the $^3P_0$ framework~\cite{LeYaouanc:1977fsz,LeYaouanc:1977gm}. We tune the parameter $\beta$ along with the coupling strength $g$ of the model. The coupling constant $g$ is fitted by choosing $\beta=0.4$ for all involved mesons, which agree with recent similar studies~\cite{Liu:2011yp}. In principle, one can choose different values of $\beta$ for initial and final mesons to be more accurate, but for simplicity we choose the same $\beta$ for all charmonium states.

\begin{table}[h]
  \renewcommand\arraystretch{2}
  \centering
\begin{tabular}{cccccccccc}
 \hline\hline
$m_c=1.50$ $\text{GeV}$ & $\beta=0.40$ $\text{GeV}$ & $g=0.80$ &$|\theta|=13\degree$\\
\hline\hline
\end{tabular}
\caption{The parameters used in our calculation.
Due to the implicit treatment of color and flavor degrees of freedom,
these factors do not show up in our calculation.}
\label{paraTab}
\end{table}

\begin{figure}[H]
  \centering
  \includegraphics[width=\textwidth]{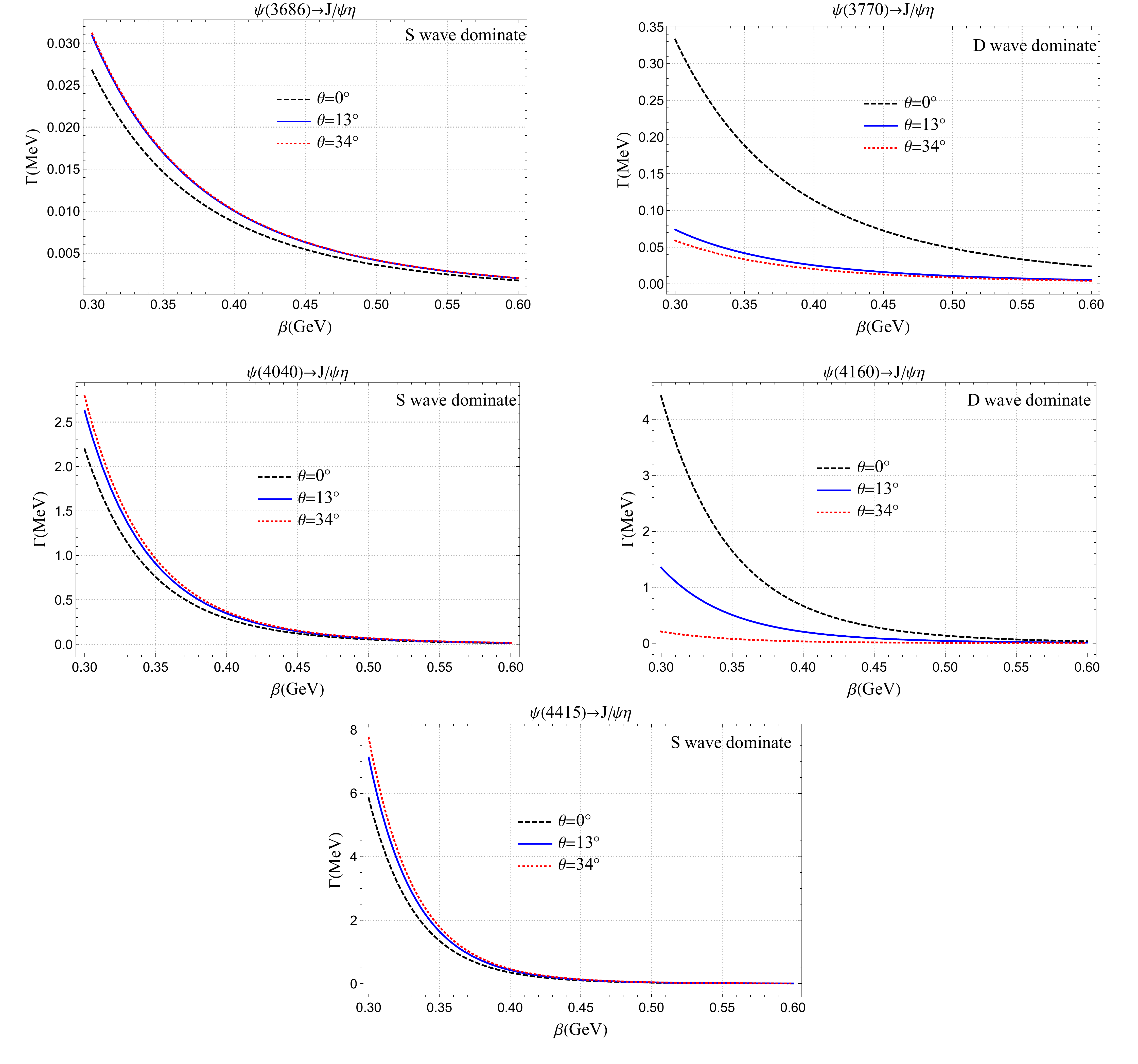}\\
  \caption{(Color online) $\beta$ dependence of decay widths of the first few $\Psi$'s into $J/\psi \eta$. The dashed black curve represents the pure $S$ and $D$ wave decay width; the solid blue and dotted red curves represent the decay width with small and large mixing angles, respectively. It is a coincidence that the decay width for $\psi(3686)\to J/\psi \eta$ is roughly the same for $\theta=13\degree$ and $\theta=34\degree$, which causes an exact overlap of the curves. It is due to the definition of the $S-D$ mixing mechanism. To confirm this argument, a rough estimate for a specific value of $\beta$ can be made from Fig.~\ref{thetaDependence}, which contains the decay width dependence on the mixing angle.}
 \label{betaDependence}
\end{figure}

\begin{figure}[H]
  \centering
  \includegraphics[width=\textwidth]{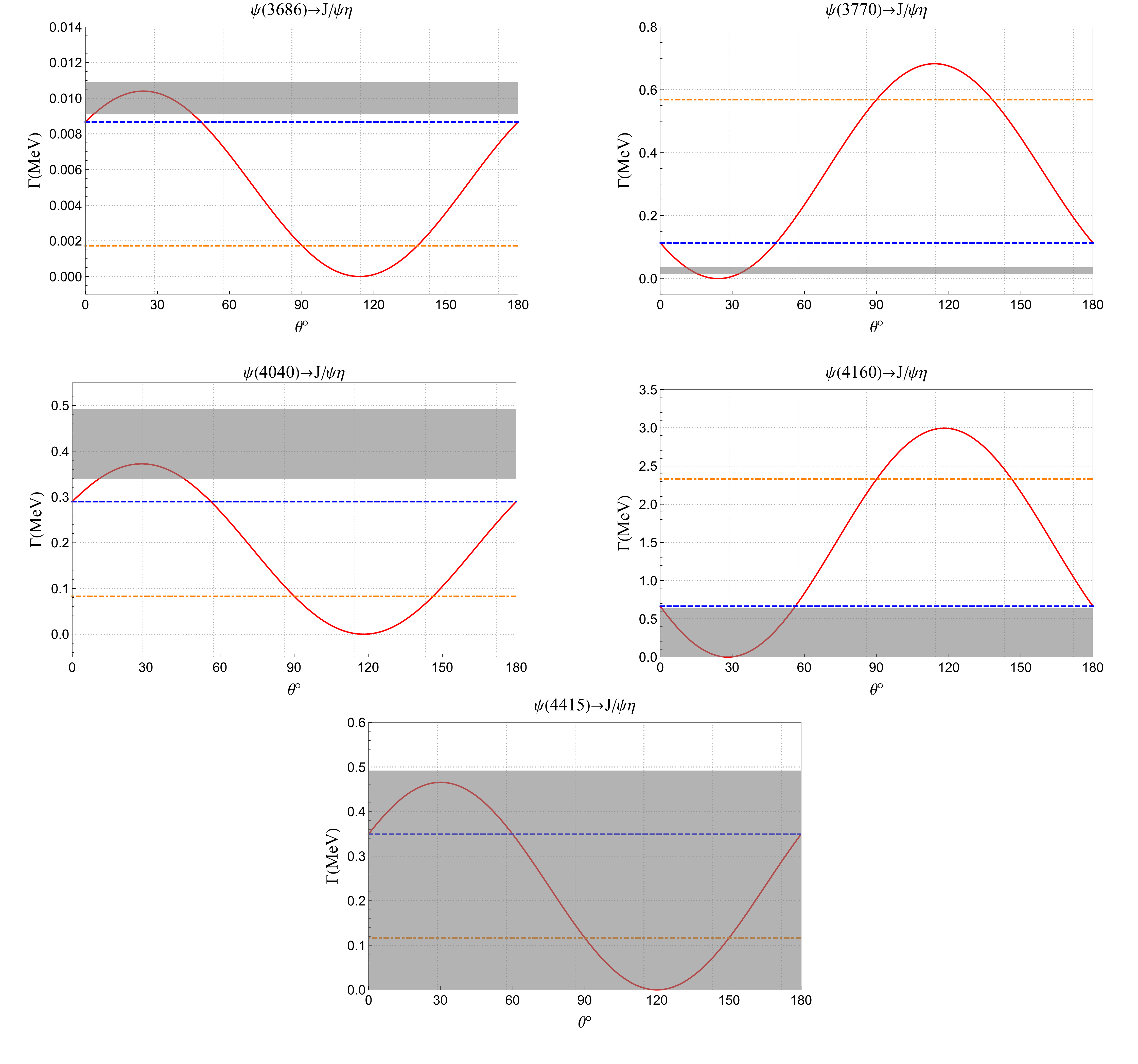}\\
  \caption{(Color online) $\theta$ dependence of the decay widths of experimentally well-established higher vector $c\bar{c}$ states decaying into $J/\psi \eta$. The dashed blue line represents the pure $S$ or $D$ wave, the dotted-dashed orange line represents $\theta=90\degree$, and the solid red curve represents the decay width as a function of the mixing angle. The gray band indicates the experimental values of $\Gamma(\Psi \to J/\psi \eta)$ with statistical errors listed in Table~\ref{eta}.}
 \label{thetaDependence}
\end{figure}

We explicitly show the $\beta$ dependence of the decay width in Fig.~\ref{betaDependence} for $\psi(3686)$, $\psi(3770)$, $\psi(4040)$, $\psi(4160)$, and $\psi(4415)$, to clarify the possible acceptable range of the $\beta$. It is worthy noting that the decay width does not change drastically around $\beta=0.4$. It is clear from Fig.~\ref{betaDependence} that one can choose any value of $\beta$ within the safe region, i.e., $0.3\sim0.6$. Our preferred value $\beta=0.4$, lies in the safe region and, hence, is perfectly adequate.

\begin{table} [H]\footnotesize
\renewcommand\arraystretch{2.5}
  \centering
\begin{tabular}{cccccc}
  \hline\hline
  State &$n ^{2S+1}L_{J}$ & $\Gamma_{\textrm{total}}$~\cite{Olive:2016xmw}&$\mathcal B(\psi \to J/\psi \eta)$~\cite{Olive:2016xmw} & $\Gamma^{\textrm{th}}_{\psi \to J/\psi \eta}$  &$\Gamma^{\textrm{exp}}_{\psi \to J/\psi \eta}$~\cite{Olive:2016xmw} \\ \hline
  $\psi(3686)$ & $2 ^3S_1$  &$0.296\pm 0.008$ & $(3.36\pm 0.05)\%$ & 0.010 & $0.010\pm0.001$ \\
  $\psi(3770)$& $1 ^3D_1 $  &$27.2\pm1.0$     & $9\pm4\times10^{-4}$ & 0.025 & $0.025\pm0.011$  \\
  $\psi(4040)$& $3 ^3S_1 $  &$80\pm 10$       &$5.2\pm 0.7\times 10^{-3}$ & 0.347 & $0.416\pm0.076$ \\
  $\psi(4160)$& $2 ^3D_1 $  &$70\pm10$        & $<8\times 10^{-3}$ &  0.204 & $<0.560\pm0.080$ \\
  $\psi(4415)$& $4 ^3S_1 $  &$62\pm20$        & $<6\times 10^{-3}$ & 0.425 & $<0.372\pm0.120$  \\
  \hline \hline
\end{tabular}
\caption{All the widths are in units of MeV. For expressing the quantum numbers, we use the spectroscopic notation $n ^{2S+1}L_{J}$, where $n=n_{r}+1$; $n_r$ is the radial quantum number; and $S$, $L$, and $J$ represent the spin, orbital, and total angular momentum of charmonia, respectively.}
\label{eta}
\end{table}

The parameters used in our calculations are listed in Table~\ref{paraTab}. Table~\ref{eta} shows the fitted results with the choice of best-fit values of the parameters of Table~\ref{paraTab}. We get quite impressive agreement with the experimental data. Although there exist only upper limits for the $\psi(4160)\to J/\psi \eta$ and $\psi(4415)\to J/\psi \eta$ decay processes, our computed decay widths for the former decay process lie within this limit, while for the latter process our predicted width is slighter larger than the central value. It is worthy noting that the experimental value of $\Gamma({\psi(4415) \to J/\psi \eta}$) has large statistical errors. Considering this error range, our prediction in this case still lies within the upper limit.

Error estimation in the theoretical model is still an open question. It became an important debate among theoretical constructs in the last few years. In general, there are no surefire prescriptions for assigning error bars to theoretical models. In the case of parameter dependence, by doing numerical analysis one can confine the model space to a \textit{physically reasonable} domain. Within this domain, there is a range of \textit{reasonable} parametrizations that can be considered as delivering a decent fit~\cite{Dobaczewski:2014jga}. Using this prescription we scan the parameters of our model and find the physically reasonable range of $\beta=0.3\sim0.6$ GeV and $|\theta|=10\degree \sim 13\degree$. For giving an idea of the uncertainties arising from model parameters, we explicitly give the decay widths in Table~\ref{error} by varying $\beta$ and $\theta$ in the described physical range.

\begin{table} [H]\footnotesize
\renewcommand\arraystretch{2}
  \centering
\begin{tabular}{c|cccc|cccc|c}
  \hline\hline
   &\multicolumn{8}{c|}{$\Gamma^{\textrm{th}}_{\psi \to J/\psi \eta}$} & \\\cline{2-9}
  State& $\beta=0.35$& $\beta=0.40$ &$\beta=0.45$ & $\beta=0.50$ &$|\theta|=10\degree$ & $|\theta|=11\degree$& $|\theta|=12\degree$ & $|\theta|=13\degree$  & $\Gamma^{\textrm{exp}}_{\psi \to J/\psi \eta}$~\cite{Olive:2016xmw}\\
  \hline
  $\psi(3686)$ & $0.017$  & $0.010$  & $0.006$ & $0.004$ &$9.782\times 10^{-3}$ & $9.865\times 10^{-3}$ & $9.942\times 10^{-3}$ & $0.010$ & $0.010\pm0.001$ \\
  $\psi(3770)$ & $0.042$  & $0.025$  & $0.016$ & $0.011$ &$0.041$ & $0.035$ & $0.030$ & $ 0.025 $ & $0.025\pm0.011$  \\
  $\psi(4040)$ & $0.906$  & $0.347$  & $0.146$ & $0.066$ &$0.336$ & $0.340$ & $0.344$ & $0.347$ & $0.416\pm0.076$ \\
  $\psi(4160)$ & $0.505$  & $0.204$  & $0.089$ & $0.041$ &$0.290$ & $0.260$ & $0.231$ & $0.204$ & $<0.560\pm0.080$ \\
  $\psi(4415)$ & $1.647$  & $0.425$  & $0.123$ & $0.039$ &$0.411$ & $0.416$ & $0.421$ & $0.425$ & $<0.372\pm0.120$ \\
  \hline \hline
\end{tabular}
\caption{All the widths are in units of MeV and rounded to $0.001$ MeV. While varying any parameter others are fixed and given in Table~\ref{paraTab}.}
\label{error}
\end{table}

\begin{figure}[H]
  \centering
  \includegraphics[width=\textwidth]{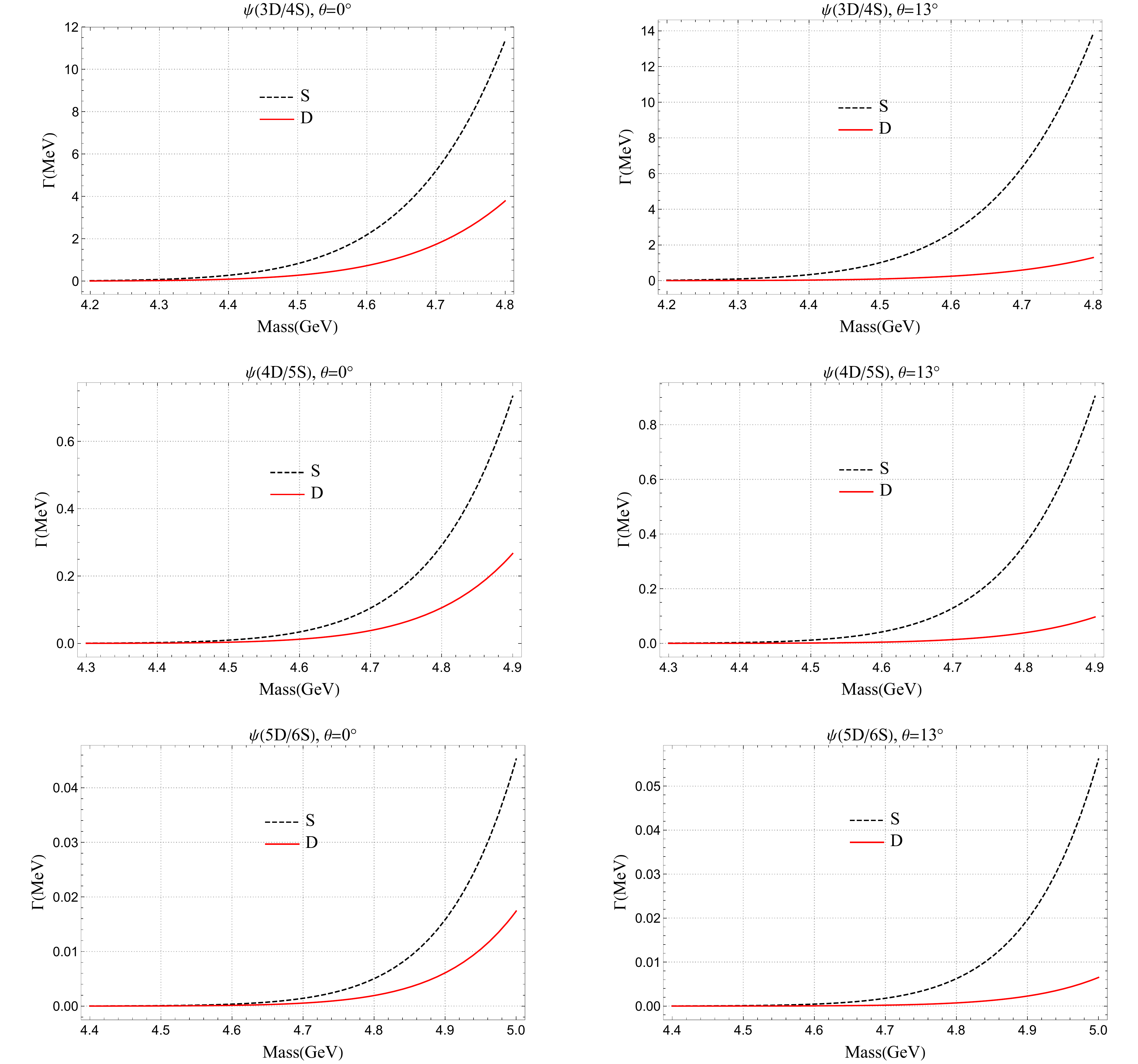}\\
  \caption{(Color online) Decay width dependence of $\Psi \to J/\psi \eta$ on initial mass. Dashed black and red solid curves represent the decay width of $S$ and $D$ wave dominate states, respectively.}
 \label{massDependence}
\end{figure}

The dependence of the decay widths on the $S-D$ mixing angle $\theta$ is very crucial to understand the behavior of this hadronic transition. We show the $\theta$ dependence of $\Gamma(\Psi\to J/\psi\eta)$ in Fig.~\ref{thetaDependence} along with a band gap, which actually represents the decay width range when we consider the corresponding $\psi$ state as pure $S$ and $D$ wave states. Due to sinusoidal behavior, small mixing angles are adequate, while large mixing angles may ruin the predictions. Although Fig.~\ref{thetaDependence} contains the predictions with a specific value of $\beta$, still it can be seen that the decay widths of the process $\psi(3686)\to J/\psi\eta$ at $\theta=13\degree$ and $\theta=34\degree$ are exactly the same, which causes an exact overlap of the two curves in Fig.~\ref{betaDependence}.

We also give the estimates of the decay widths of $\psi(3^3D_1)$, $\psi(4^3S_1)$, $\psi(4^3D_1)$, $\psi(5^3S_1)$, $\psi(5^3D_1)$, and $\psi(6^3S_1)$ states decaying to $J/\psi \eta$ in Fig.~\ref{massDependence}, which provide useful information to search and understand the missing higher vector states. Due to the fact that these higher states are poorly understood experimentally, we are not able to predict the widths exactly. To plot the decay width as a function of the mass of the corresponding higher $c\bar{c}$ vector state, we consider the mass range based on serval quark model predictions of the mass spectra listed in Table~\ref{masses}.

\begin{table}[H]\footnotesize
\renewcommand\arraystretch{2}
  \centering
\begin{tabular}{ccccccc}
  \hline\hline
  State & $J^{PC}$& Screened~\cite{Li:2009zu}& BGS-NR~\cite{Ding:2007rg}& BGS-Rel.~\cite{Sultan:2014oua} & CQM~\cite{Segovia:2008zz} &RS~\cite{Badalian:2008dv} \\
  \hline
  $\psi(4^3S_1)$ & $1^{--}$ & $4273$ & 4406~\footnotemark[2]  & 4356 & 4389 & 4420\\
  $\psi(3^3D_1)$ & $1^{--}$ & 4317 & 4455  & 4470 & 4426 & 4470\\
  $\psi(5^3S_1)$ & $1^{--}$ & 4463 & 4704 & 4661 & 4614 & 4655\\
  $\psi(4^3D_1)$ & $1^{--}$ & $-$ & 4770  & 4735 & 4641 & 4700\\
  $\psi(6^3S_1)$ & $1^{--}$ & 4608 & 4977 & 4912 & 4791 & 4815\\
  $\psi(5^3D_1)$ & $1^{--}$ & $-$ & $-$  & 4976 & 4810 & $-$\\
  \hline\hline
\end{tabular}
\caption{Quark model predictions of mass spectra for higher vector charmonia. Reference~\cite{Li:2009zu} predicts mass in a nonrelativistic screened potential model, which incorporates the color screening effects due to the creation of a light $q\bar{q}$ pair within the heavy $Q\bar{Q}$. Reference~\cite{Ding:2007rg} contains the predictions of the nonrelativistic effective $Q\bar{Q}$ potential as calculated by Barnes-Godfrey-Swanson (BGS)~\cite{Barnes:2005pb}. Predictions of BGS potential with relativistic corrections to mass (calculated by using leading-order perturbation theory) are taken from Ref.~\cite{Sultan:2014oua}. Predications of the constituent quark model (CQM)~\cite{Segovia:2008zz} and Salpeter equation with relativistic string (RS) Hamiltonian~\cite{Badalian:2008dv} are also listed. $``-"$ indicates that the prediction is not available.}
\label{masses}
\end{table}
\footnotetext[2]{This value is taken from Ref.~\cite{Barnes:2005pb}.}

We include estimates of pure higher $S$ and $D$ wave states along with the predictions with small $S-D$ mixing. In all considered cases, the decay width of the $D$ wave state is smaller than that of the corresponding $S$ wave one. Given that the $D$ wave interferes destructively with the $S$ wave, its decay width is suppressed significantly with the use of a small mixing angle. After a particular mass value, the decay width of both $S$ and $D$ wave states becomes very sensitive to the initial mass. For example in the $\psi(3D/4S)$ case, the decay width rises exceptionally after $4.5$ GeV. This critical mass value for other higher states can be observed from Fig.~\ref{massDependence}. With reference to the observed order of the $\eta$ emission rate, these critical mass values might provide the upper limits on the mass of the corresponding higher vector charmonium.

\subsection{Predictions for $\psi(nS/(n-1)D) \to h_{c}(1P) \eta$ with $(n=3,4,5)$}
\label{hc}

The hidden-charm $\eta$ decay of $J^{PC}=1^{--}$ higher charmonia into the lowest $P$ wave $c\bar{c}$ meson, i.e., $h_{c}(1^1P_1)$, is also possible. The threshold for this decay process is $4073$ MeV. So, the first vector state which can decay into $h_{c}(1P) \eta$ is $\psi(4160)$. There exists an experimental evidence for $e^{+}e^{-} \to h_{c}(1P)\eta$ around $4170$ MeV mass by the CLEO Collaboration~\cite{CLEO:2011aa}. Their reported measurement on the branching fraction $\mathcal B (\psi(4160)\to h_{c}(1P)\eta)$ is $< 2\times 10^{-3}$ with a $90\%$ confidence level. The HQSS violating transition $\psi \to h_{c}(1P) \eta$ requires the spin flip to be significantly suppressed relative to the corresponding heavy quark spin-conserving transitions like $\Psi \to J/\psi \pi \pi$~\cite{Voloshin:2012dk}. The observed ratio $\Gamma(\Psi \to J/\psi \eta)/\Gamma(\Psi \to J/\psi \pi \pi)$ is fully consistent with the earlier theoretical predictions~\cite{Voloshin:2007dx}. It has been argued in~\cite{Guo:2010ak,Guo:2010zk} that the coupled-channel effects due to intermediate charmed mesons for these transitions are quite small.

\begin{figure}[H]
  \centering
  \includegraphics[width=\textwidth]{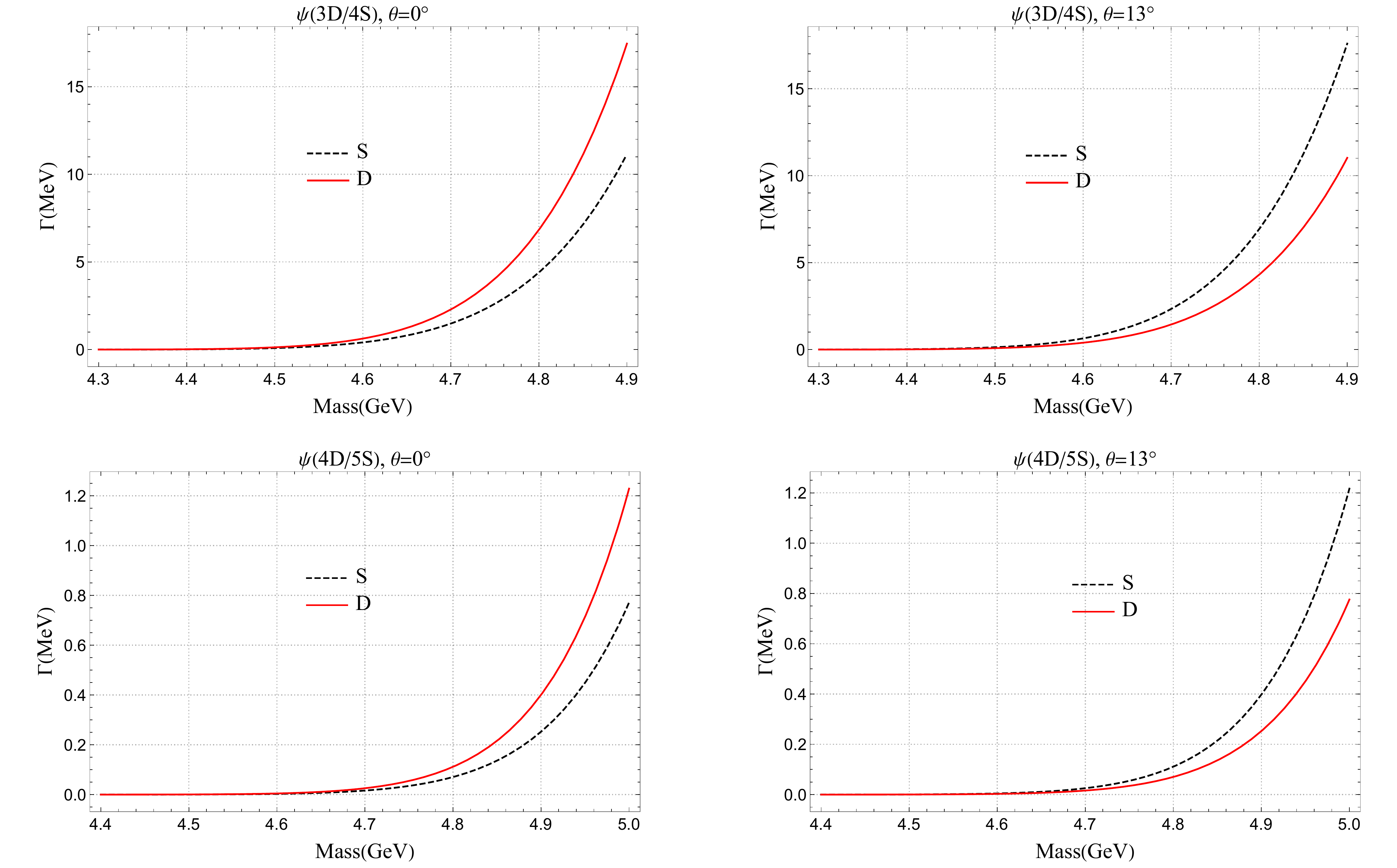}\\
  \caption{(Color online) Decay width dependence on the initial mass for a few $\Psi$'s into $h_{c}(1P)\eta$. Dashed black and solid red curves represent the decay width of $S$ and $D$ wave dominate states, respectively.}
 \label{hcwidth}
\end{figure}

Among well-known higher vector charmonia, only $\psi(4160)$ and $\psi(4415)$ have enough phase space to decay into $h_{c}(1P) \eta$. Table~\ref{hc1P} contains our predictions for these states. We predict the decay width of $\psi(4415)\to h_{c}(1P) \eta$ with the same order of magnitude as $\psi(4160)$ to the similar final state:
\be
\frac{\Gamma(\psi(4160)\to h_{c}(1P) \eta)}{\Gamma(\psi(4160)\to J/\psi \eta)}=7.887\times 10^{-2},
\label{hc2D}
\ee
\be
\frac{\Gamma(\psi(4415)\to h_{c}(1P) \eta)}{\Gamma(\psi(4415)\to J/\psi \eta)}=6.736\times 10^{-2}.
\label{hc4S}
\ee
It is not easy to give the estimates of the decay width for $\psi(3D)$, $\psi(4D)$, and $\psi(5S)$ or higher ones because these states have not been experimentally well established up to now and hence, their masses are unknown. We give the initial mass dependence of the decay width of the $\Psi \to h_{c}(1P) \eta$ transition of these higher vector states in Fig.~\ref{hcwidth}, both for the pure $S$ and $D$ wave and for the standard $S-D$ mixing case.

\begin{table} [H]\footnotesize
\renewcommand\arraystretch{2.5}
  \centering
\begin{tabular}{cccccc}
  \hline\hline
  State &$n ^{2S+1}L_{J}$ & $\Gamma_{\textrm{total}}$~\cite{Olive:2016xmw}&$\mathcal B(\psi \to h_{c}(1P) \eta)$~\cite{CLEO:2011aa} & $\Gamma^{\textrm{th}}_{\psi \to h_{c}(1P) \eta}$  &$\Gamma^{\textrm{exp}}_{\psi \to h_{c}(1P) \eta}$~\cite{CLEO:2011aa} \\ \hline

  $\psi(4160)$& $2 ^3D_1 $  &$70\pm10$        & $<2\times 10^{-3}$ &  $  1.609\times10^{-2}$ & $<0.140\pm0.020$ \\
  $\psi(4415)$& $4 ^3S_1 $  &$62\pm20$        & $-$ & $ 2.863\times10^{-2}$ & $-$  \\
  \hline \hline
\end{tabular}
\caption{All the widths are in units of MeV. $``-"$ indicates that the experimental data are not available.}
\label{hc1P}
\end{table}

For $\Psi \to h_{c}(1P)\eta$, the decay width of the pure $D$ wave is larger than that of the corresponding pure $S$ wave. It is due to that the overlap of the wave function of the $D$ wave with $1P$ is larger than the corresponding $S$ wave. The $D$ wave interferes destructively with the $S$ wave; the decay width of $D$ wave, in this case, is not suppressed significantly as we have observed for the $\Psi \to J/\psi\eta$ channel. For the $\Psi \to J/\psi \eta$ case, the reason why the decay width of the pure $D$ wave is smaller is that the overlap of the $D$ wave with $1S$ is smaller than the corresponding $S$ wave state. The swapping of the curves can be seen in Fig.~\ref{hcwidth} with the increase of the mixing angle. The constructive interference of the $S$ wave with the $D$ wave causes a significant increase in the decay width of the $S$ wave dominate state.

\subsection{$Y(4360)$, $Y(4390)$ and $Y(4660)$ Assignments}
\label{assignments}

Just after the experimental observation of $Y(4360)$ and $Y(4660)$ in the initial state radiation (ISR) $e^+e^-\to\gamma_{\textrm{ISR}} \pi^+ \pi^- \psi(2S)$ process at Belle~\cite{Wang:2007ea}, many theoretical studies were carried out to incorporate these states into conventional and exotic $c\bar{c}$ spectra (for an overview, see the discussion in Sec. 4.8 of the recent review~\cite{Chen:2016qju}). Among these configurations, there exist a couple of interpretations by considering $Y(4360)$ and $Y(4660)$ as canonical $3 ^3D_1$~\cite{Ding:2007rg} and $5 ^3S_1$ charmonium~\cite{Ding:2007rg,Segovia:2008zz}, respectively. Predictions were also made for the dielectric widths, E1 and M1 transitions, and open flavor strong decays. In the screened potential model, the state $Y(4360)$ is also interpreted as $\psi(3 ^3D_1)$, while $Y(4660)$ was considered as $\psi(6 ^3S_1)$~\cite{Li:2009zu}. Assuming $Y(4360)$ as $3 ^3D_1$ and $Y(4660)$ as $5 ^5S_1$ dominate, their dielectric decay widths can be reproduced to get agreement with experimental data by introducing the large $S-D$ mixing~\cite{Badalian:2008dv}.

The predicted mass of $\psi(3^3D_1)$ in the BGS and RS potential model (as shown in Table~\ref{masses}) is somewhat larger than the experimental mass of $Y(4360)$. However, we notice that the mass predictions of various potential models for the higher $c\bar{c}$ states may differ by $10\sim100$ MeV~\cite{Eichten:2007qx,Swanson:2006st}. Therefore, it is not irrational to treat $Y(4360)$ as $\psi(3D)$ dominant with a small component of $\psi(4S)$, exactly in the same way as for the lower $D$ waves in Sec.~\ref{mixing}. Mass predictions for $\psi(5 ^3S_1)$ of various potential models do not differ much from $Y(4660)$ except the one produced by the screened potential model. Hence, it is also not irrational to treat $Y(4660)$ as $\psi(5S)$ dominant having a small component of $\psi(4D)$.

Very recently, the BESIII Collaboration announced the observation of a new resonant structure, $Y(4390)$ in the $e^+e^-\to \pi^+ \pi^- h_c$ process~\cite{BESIII:2016adj}. Its measured mass and total decay width are $(4391.6 \pm 6.3 \pm 1.0)$ MeV and $(139.5 \pm 16.1 \pm0.6)$, respectively. The width of $Y(4390)$ seems to be broader. It might be possible that this higher mass region could be described either by a $Y(4360)$ resonance, or by a phase-space background as has already been noticed in the case of $Y(4290)$~\cite{Esposito:2016noz}. Since the reported quantum numbers of this state are $J^{PC}=1^{--}$ and it lies in the mass region of $3D$ charmonium, hence, this new resonance may also be considered as a candidate for $\psi(3D)$. As an estimate, we give our predictions of its $\eta$ transition branching fraction in Table~\ref{assign} by assigning $3 ^3D_1$.

One may argue that the vector state $Y(4260)$ can also be assigned as a $3 ^3D_1$ or $4 ^3S_1$ charmonium state. As listed in Table~\ref{masses}, in various potential models, $\psi(3 ^3D_1)$ and $\psi(4 ^3S_1)$ are predicted between $4.3\sim4.5$ GeV. Therefore, with reference to the observed mass of $Y(4260)$, it lies below the $\psi(4S)\sim \psi(3D)$ mass region. The value of its $\Gamma_{e^{+} e^{-}}$ is two to three orders of magnitude smaller than those
of the well-established conventional vector charmonia~\cite{Lebed:2016hpi}. It is nearly impossible to accommodate $Y(4260)$ as a conventional charmonium state. Many theoretical constructs consider $Y(4260)$ as an exotic charmonium~\cite{Esposito:2016noz}, for example, as a $\bar{D}D_{1}(2420)$ molecule with a binding energy of 29 MeV~\cite{Wang:2013cya}, or as a heavy hybrid meson with a gluonic excitation of about 1 GeV higher than the lowest $\eta_{c}$ and $J/\psi$~\cite{Chen:2016qju,Lebed:2016hpi}.

Despite the fact that these $Y$ states do not decay into open-charm channels, it would be interesting to study their hidden-charm strong decays. By assuming $Y(4360)$ and $Y(4660)$ as $\psi(3^3D_1)$ and $\psi(5^3S_1)$ dominant states, respectively, we give our predictions for $Y(4360)\to J/\psi \eta$ and $Y(4660)\to J/\psi \eta$, which might be helpful to understand the properties of these vector states. Because only experimental upper limits~\cite{Wang:2012bgc} exists for the product of the branching fraction $\mathcal B(Y\to J/\psi \eta)$ and $\Gamma_{e^+ e^-}(Y)$ for $Y(4360)$ and $Y(4660)$, we need to know the dielectric decay width of these states. There exist few theoretical predictions for $\Gamma_{e^+ e^-}$ for these states as summarized in Table~\ref{dielectric}.

\begin{table}[H]
\renewcommand\arraystretch{2}
  \centering
\begin{tabular}{ccccc}
  \hline\hline
  Initial state & $n ^{2S+1}L_J$ & Mass (MeV)& $\Gamma_{e^+ e^-} (\textrm{keV})$ \\
  \hline
  $\Gamma (Y(4360)\to e^+ e^-)$ & $3 ^3D_1$ & 4361 & 0.87~\cite{Ding:2007rg} \\
   && 4455 & 0.83~\cite{Ding:2007rg}  \\
   && 4426~\footnotemark[3] & 0.33~\cite{Segovia:2008zz}  \\
   && 4470 & 0.06~\cite{Badalian:2008dv}  \\
   && 4470 & 0.63~\footnotemark[4]~\cite{Badalian:2008dv}\\
  $\Gamma (Y(4660)\to e^+ e^-)$ &$5 ^3S_1$ & 4664 & 1.34~\cite{Ding:2007rg} \\
   && 4704 & 1.32~\cite{Ding:2007rg} \\
    && 4614 & 0.57~\cite{Segovia:2008zz} \\
   && 4655 & 0.73~\cite{Badalian:2008dv} \\
    && 4655  & 0.39~\footnotemark[4]~\cite{Badalian:2008dv} \\
  \hline\hline
\end{tabular}
\caption{$e^+ e^-$ dileptonic partial decay widths for $Y(4360)$ and $Y(4660)$ states.}
\label{dielectric}
\end{table}
\footnotetext[3]{Reference~\cite{Segovia:2008zz} assigned $3 ^3D_1$ to $\psi(4415)$ instead of $Y(4360)$, while our concern here is just having a comparison among predictions of the dielectric decay width of $3 ^3D_1$ with different input masses.}
\footnotetext[4]{This prediction contains the admixture of $4 ^3S_1$ for $Y(4360)$ and $4 ^3D_1$ for $Y(4660)$ with mixing angle $\theta=34\degree$.}

As shown in Table~\ref{assign}, the only available experimental information is $\mathcal B(Y\to J/\psi \eta)\cdot \Gamma_{e^+ e^-}^{Y}$ for these $Y$ states. Therefore, we need $\Gamma_{e^+ e^-}$ to compute $\Gamma(Y\to J/\psi \eta)$. Table~\ref{dielectric} contains different theoretical predictions for $\Gamma_{e^+ e^-}$ of these $Y$ states both for pure $S$ and $D$ waves, and with the large mixing ($\theta=34\degree$). For $Y(4360)$, we take the average value of $\Gamma_{e^+ e^-}=0.523$ keV from Table~\ref{dielectric} by considering it as pure $3D$. The experimental upper limit for $\Gamma (Y(4360)\to J/\psi \eta)$ is given in Table~\ref{assign}. To give a comparison with Ref.~\cite{Badalian:2008dv}, we also compute the upper limit of $\Gamma (Y(4360)\to J/\psi \eta)$ with large $S-D$ mixing, i.e., $\theta=34\degree$. In all three cases, pure $3D$, small, and large mixing, our predictions are in agreement with the experimental measurements. We conclude that $Y(4360)$ could be considered as a potential candidate for dominant $3 ^3D_1$ charmonium state.

\begin{table} [H]\footnotesize
\renewcommand\arraystretch{1.5}
  \centering
\begin{tabular}{cccc|ccc|ccc}
  \hline\hline
   & & & &\multicolumn{3}{c|}{$\Gamma^{\textrm{th}}_{Y \to J/\psi \eta}$} & \multicolumn{2}{c}{$\Gamma^{\textrm{exp}}_{Y \to J/\psi \eta}$} & \\\cline{5-10}
  State& $n ^{2S+1}L_{J}$ &$\Gamma_{\textrm{total}}$ &$\mathcal B(Y\to J/\psi \eta)$ &$\theta=0\degree$ & $\theta=13\degree$& $\theta=34\degree$ & $\theta=0\degree$  & $\theta=34\degree$ \\
  \hline
  $Y(4360)$ & $3 ^3D_1$  & $ 74\pm 18$~\cite{Olive:2016xmw}  & $\frac{6.8}{\Gamma_{e^+ e^-}}$~\cite{Wang:2012bgc} &  0.047 & 0.016& $1.0\times 10^{-3}$ & $<0.963$ &$<0.799$
  \\
  $Y(4390)$& $3 ^3D_1 $  & $139.5\pm16.1$~\cite{BESIII:2016adj} & $-$ &$0.083$ & $0.028$ &$ 1.6 \times 10^{-3}$  & $-$  & $-$ &  \\
  $Y(4660)$& $5 ^3S_1 $  & $48\pm 15$~~\cite{Olive:2016xmw}  & $\frac{0.94}{\Gamma_{e^+ e^-}}$~\cite{Wang:2012bgc} & 0.057 & 0.070  &0.077&  $<0.046$ & $<0.116$\\
  \hline \hline
\end{tabular}
\caption{Predictions for $\Gamma(Y \to J/\psi \eta)$ for the $Y(4360)$, $Y(4390)$, and $Y(4660)$ states. $``-"$ indicates that the experimental data are not available. All the widths are in units of MeV.}
\label{assign}
\end{table}

For $Y(4660)$, we also include the predictions for pure $5S$ and the mixed case. For the pure $5 ^3S_1$ state, we take the average value of $\Gamma_{e^+ e^-}=0.99$ keV from Table~\ref{dielectric} and list the experimental upper limit along with our prediction for $\Gamma(Y(4660)\to J/\psi \eta)$ in Table~\ref{assign}. In the large $S-D$ mixing case ($\theta=34\degree$), the dielectric decay width $=0.39$ keV~\cite{Badalian:2008dv} allows us to give an upper limit on $\Gamma (Y(4660)\to J/\psi \eta)$. Our predicted value, in this case, is within this upper limit as shown in Table~\ref{assign}. In all cases, our predictions agree with the experimental data. Hence, our results are consistent with the experimental data and the state $Y(4660)$ can be considered as $\psi(5 ^3S_1)$ dominant with a small $\psi(4 ^3D_1)$ component.

In the case of $Y(4390)$, for the sake of completeness, we give our predictions of its hidden charm $\eta$ decay, with and without $S-D$ mixing. To identify this state, measurements on its hadronic branching fraction are required. We think that these estimates might be useful to clarify the picture of these vector states and give some references to search for the missing higher $S$ and $D$ wave vector charmonia.

\section{Summary}
\label{summary}

A model to create a light meson for heavy quarkonium transition is proposed. This model is used to study the decays of higher vector charmonia into $J/\psi \eta$ and $h_{c}(1P)\eta$. Computed decay widths are in excellent agreement with experimental data. The ratio $\Gamma(\Psi\to h_{c}(1P)\eta)/\Gamma(\Psi\to J/\psi\eta)$ is predicted for $\psi(4160)$ and $\psi(4415)$. The initial state's mass dependence of $\Gamma(\Psi\to h_{c}(1P)\eta, J/\psi\eta)$ for higher vector charmonium is given. We suggest that the ongoing (Belle and BESIII) and forthcoming ($\bar{\textrm{P}}$ANDA and BelleII) experiments look for suggested unobserved decay channels. We also give the estimates of $\eta$ transition branching fractions for $Y(4360)$, $Y(4390)$, and $Y(4660)$ by assuming them as $c\bar{c}$ bound states with quantum numbers $3 ^3D_1$, $3 ^3D_1$, and $5 ^3S_1$, respectively. Our predictions reflect that the state $Y(4360)$ can be considered as a potential candidate for the $3 ^3D_1$ charmonium state. Assuming $Y(4660)$ to be $5 ^3S_1$, the predictions are consistent with the experimental upper limit. For a broader $Y(4390)$ state, the update on its hadronic branching fraction from BESIII is eagerly awaited. We hope that our predictions might provide useful references to determine the properties of higher charmonium states in ongoing and forthcoming experiments.

\section*{Acknowledgements}

The authors are grateful to Feng-Kun Guo, Qiang Zhao, Martin Cleven, Eric Swanson and Ahmed Ali for useful discussions and suggestions. We are indebted to Shan-Gui Zhou for providing us Ref.~\cite{Dobaczewski:2014jga} during the revision of this manuscript. M. N. A. gratefully acknowledges the hospitality at the DESY theory division, where part of this work was carried out. This work is supported by the National Natural Science Foundation of China under Grant No.~11261130311 (CRC110 by DFG and NSFC). M. N. A. received support from the CAS-TWAS President's Fellowship for International Ph.D. Students.

\subsubsection*{Note added in proof}

After the revision of this manuscript, BESIII published an evidence of $e^+e^- \to \eta h_c$ at center-of-mass energy $\sqrt{s}=4.358$ GeV~\cite{Ablikim:2017nkn}. Along with their earlier measurement~\cite{Ablikim:2015xhk}, this evidence will help to delve into the $Y(4360)$ through its HQSS violating hadronic transitions.
\newpage
\begin{appendix}

\section{Free-Quark Amplitude}
\label{app:amp}
To evaluate the matrix element of $A\to BC$ decay, we need spin matrix elements. At the quark level, these matrices involve the matrix elements of the Dirac bilinear (with $\Gamma=i{\gamma}^5$ and $I$ in our case) and Pauli matrix elements. Matrix elements of the Dirac bilinear in the nonrelativistic limit can be expressed as
\begin{equation}
\lim_{q \to 0}\bar{u}_{q' s'}\Gamma u_{q s}=
  \begin{cases}
    \delta_{ss'} & \Gamma=I \\
    \frac{i}{2m_q}\langle s' \vert \vec{\sigma} \vert s \rangle \cdot (\vec{q}-\vec{q}') &  \Gamma=i\gamma^5
   \end{cases}
\end{equation}
\begin{equation}
\lim_{q \to 0}\bar{v}_{\bar{q} \bar{s}}\Gamma v_{\bar{q}' \bar{s}'}=
  \begin{cases}
    -\delta_{\bar{s}\bar{s}'} & \Gamma=I \\
    \frac{i}{2m_q}\langle \bar{s} \vert \vec{\sigma} \vert \bar{s}' \rangle \cdot (\vec{\bar{q}}-\vec{\bar{q}}') &  \Gamma=i\gamma^5 .
   \end{cases}
\end{equation}
We have already shown that it is easy to handle the wave function overlap integration in a Cartesian basis; therefore, we express the elements of Pauli spinors in terms of Cartesian basis vectors as
\begin{equation}
\begin{cases}
    \langle \downarrow \vert \vec{\sigma} \vert \uparrow \rangle = (\hat{x}+i\hat{y})\\
    \langle \uparrow \vert \vec{\sigma} \vert \downarrow \rangle = (\hat{x}-i\hat{y})\\
    \langle \uparrow \vert \vec{\sigma} \vert \uparrow \rangle = - \langle \downarrow \vert \vec{\sigma} \vert \downarrow \rangle = \hat{z}
\end{cases}
\end{equation}
For antiquark case, these relations are
\begin{equation}
\begin{cases}
    \langle \bar{\downarrow} \vert \vec{\sigma} \vert \bar{\uparrow }\rangle = -(\hat{x}-i\hat{y})\\
    \langle \bar{\uparrow} \vert \vec{\sigma} \vert \bar{\downarrow }\rangle = -(\hat{x}+i\hat{y})\\
    \langle \bar{\uparrow }\vert \vec{\sigma} \vert \bar{\uparrow} \rangle = - \langle \bar{\downarrow }\vert \vec{\sigma} \vert \bar{\downarrow} \rangle = - \hat{z}
\end{cases}
\end{equation}
From Fig.~\ref{fyn2}, one can get the following relation of momentum conservation by considering the initial state in the center-of-mass reference frame:
\be
\begin{cases}
p_2=-p_1\\
p'_1=p_1-k\\
P_B=-k .
\end{cases}
\ee
The meson's  space wave function can be written as
\be
\begin{cases}
\phi_A\{\mu_A(\frac{p_1}{m_1}-\frac{p_2}{m_2})\}=\phi_A(p_1)\\
\phi_B\{\mu_B(\frac{p'_1}{m_1}-\frac{p_2}{m_2})\}=\phi_B(p_1-\frac{m_1}{m_1+m_2}P_B)\\
\end{cases}
\ee
where the $\mu_i,i=(A,B)$  are the reduced masses of the constituent quarks of the mesons $A$ and $B$, respectively. It should be noted that all the momenta should and can be expressed by the integration variable and the momentum of the $B$ meson, viz. $p_1$ and $P_B$ in this work. The free-quark amplitude for Fig.~\ref{fyn2} is
\be
i\mathcal{M}_0=[\bar{u}_{q's'}(i\gamma^5)u_{qs}] [\bar{v}_{\bar{q}\bar{s}} I v_{\bar{q}'\bar{s}'}],
\ee
\be
i\mathcal{M}_0=\frac{i}{2m_c} \langle 1' \vert \vec{\sigma} \vert 1 \rangle \cdot (\vec{p}_1-\vec{p'}_{1}) \cdot \langle 2 \vert \delta_{ss'} \vert 2' \rangle,
\ee
where $m_c$ is the mass of the charm quark. Collecting all the pieces together, the full amplitude becomes
\be
\helicityAmp =g \frac{i}{2m_c} \int d^3 p_1
\phi_A(\vec{p}_1) \phi_B^*(\vec{p}_1-x_B \vec{P}_B) \langle 1' \vert \vec{\sigma} \vert 1 \rangle \cdot (\vec{p}_1-\vec{p'}_{1}) \cdot \langle 2 \vert \delta_{ss'} \vert 2' \rangle .
\label{fianl:amp}
\ee
This is the nonrelativistic approximation of Eq.~(\ref{amp}). The momentum $P_B$ can be computed by using Eq.(\ref{mom}).
\end{appendix}

\providecommand{\href}[2]{#2}\begingroup\raggedright\endgroup
\end{document}